\documentclass[modern, a4paper, times]{aastex63}
\usepackage[english]{babel}

\usepackage{amsmath}
\usepackage{graphicx}
\usepackage{natbib}

\newcommand{\kms}{~km~s$^{-1}$}
\newcommand{\obs}{_{\text{obs}}}
\newcommand{\HP}{\ensuremath{_{\text{HP}}}}
\newcommand{\start}{\ensuremath{_{\text{in}}}}
\newcommand{\stopp}{\ensuremath{_{\text{out}}}}
\newcommand{\myvec}[1]{\ensuremath{\boldsymbol{{#1}}}}

\begin{document}
\title{Time delay between outer heliosheath crossing and observation of interstellar neutral atoms}
\shorttitle{Times of flight of ISN atoms}
\shortauthors{M. Bzowski \& M.A. Kubiak}

\correspondingauthor{M. Bzowski}
\email{bzowski@cbk.waw.pl}

\author[0000-0003-3957-2359]{M. Bzowski}
\affiliation{Space Research Centre, Polish Academy of Sciences (CBK PAN),\\
Bartycka 18A, 00-716 Warsaw, Poland}

\author[0000-0002-5204-9645]{M.A. Kubiak}
\affiliation{Space Research Centre, Polish Academy of Sciences (CBK PAN),\\
Bartycka 18A, 00-716 Warsaw, Poland}
\keywords{Heliosphere (711), Heliosheath (710),  Heliopause (707), Astrosphere interstellar medium interactions (106), Stellar wind bubbles (1635)}

\begin{abstract}
In situ measurements of the heliospheric particle populations by the Voyager spacecraft can only be put in an appropriate context with remote-sensing observations of energetic and interstellar neutral atoms (ENAs and ISN, respectively) at 1 au when the time delay between the production and the observation times is taken into account. ENA times of flight from the production regions in the heliosheath are relatively easy to estimate because these atoms follow almost constant speed, force-free trajectories. For the ISN populations, dynamical and ballistic selection effects are important, and times of flight are much longer. We estimate these times for ISN He and H atoms observed by IBEX and in the future by IMAP using the WTPM model with synthesis method. We show that for the primary population atoms, the times of flight are on the order of three solar cycle periods, with a spread equivalent to one solar cycle. For the secondary populations, the times of flight are on the order of ten solar cycle periods, and during the past ten years of observations, IBEX has been collecting secondary He atoms produced in the OHS during almost entire 19th century. ISN atoms penetrating the heliopause at the time of Voyager crossing will become gradually visible about 2027, during the planned IMAP observations. Hypothetical variations in the ISN flow in the Local Interstellar Medium are currently not detectable. Nevertheless, we expect steady-state heliosphere models used with appropriately averaged solar wind parameters to be suitable for understanding the ISN observations. 
\end{abstract}

\section{Introduction}
\label{sec:intro}
\noindent
The heliosphere is formed due to interaction of the solar wind with the local interstellar matter. This interaction has been studied using both in situ and remote-sensing observation techniques. Understanding the processes at the heliospheric boundary and of the conditions outside the heliosphere is typically done by fitting parameters used in models of the heliosphere to various observations. The observables include, e.g., the distances of crossing the termination shock and the heliopause by the Voyager spacecraft, the sky distribution of the flux of energetic neutral atoms (ENAs), the size and the center location of the IBEX ribbon of enhanced ENA emission, the flux of interstellar neutral (ISN) gas directly sampled at 1~au, and the sky distribution of the heliospheric backscattered Lyman-$\alpha$ glow.  Application of state of the art models of the heliosphere for determination of various subsets of the Local Interstellar Medium (LISM) parameters based on various subsets available data resulted in significantly different estimates of numerical values of these parameters (e.g., \citet{zirnstein_etal:16b} and \citet{bzowski_etal:19a} on one hand and \citet{izmodenov_alexashov:20a} on the other hand).

Solar wind varies with time quasi-periodically at a time scale of the period of the solar activity cycle $\sim 10.7$~years and secularly, with the strength of the activity maximum changing from one solar cycle to another. This results in variations of the solar wind dynamic pressure and other parameters  with time. Even assuming that the interstellar matter around the Sun is homogeneous at spatial scales of a thousand au, the variation of the solar wind implies that the size of the heliosphere, the shape and distance to the solar wind termination shock, and to the heliopause vary. Also the plasma flow around the heliosphere in the outer heliosheath (OHS), as well as that in the inner heliosheath between the termination shock and the heliopause vary. Consequently, also populations of particles used as carriers of information on the processes operating in the interaction region, like ENAs, energetic ions, and the secondary population of ISN gas created in the OHS vary with time. 

Some of these populations, as well as the derivative populations created in the inner heliosphere due to interactions with the solar output, also vary in time. Because of all these variations, a priori the global heliosphere and the individual heliospheric populations should be studied using time-dependent models. Such models have been developed: see, e.g., \citet{izmodenov_malama:04a, pogorelov_etal:09c, izmodenov_alexashov:15a}. However, an open question remains what is the delay between the creation and the detection of the particle population observed nowadays at 1 au, and if the available measurements of solar activity-related phenomena reach sufficiently backwards in time to enable a realistic modeling of the heliospheric interface at the time of creation of presently-observed populations.  

\begin{figure} 
\centering 
\includegraphics[width=0.49\textwidth]{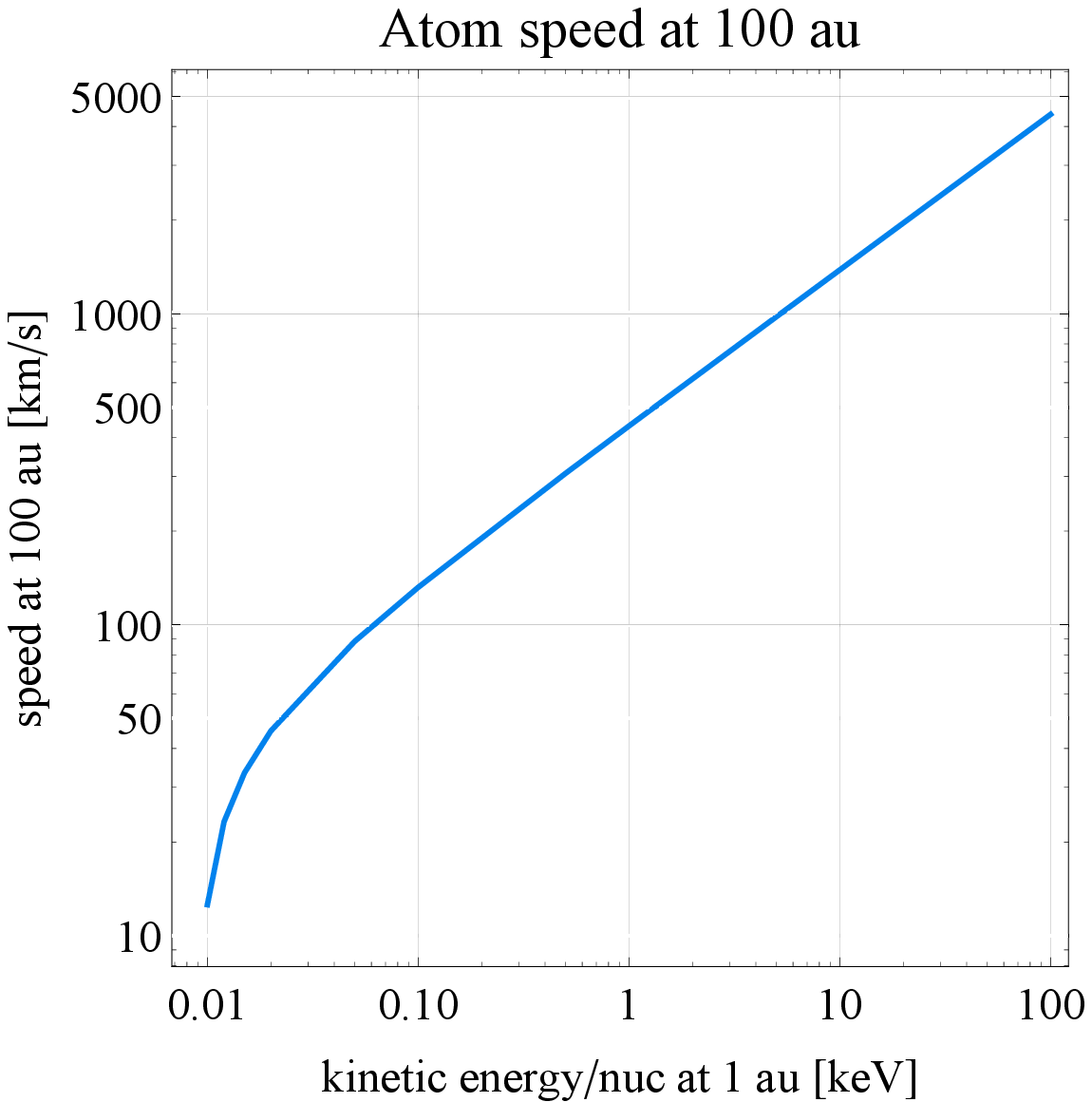} 

\includegraphics[width=0.49\textwidth]{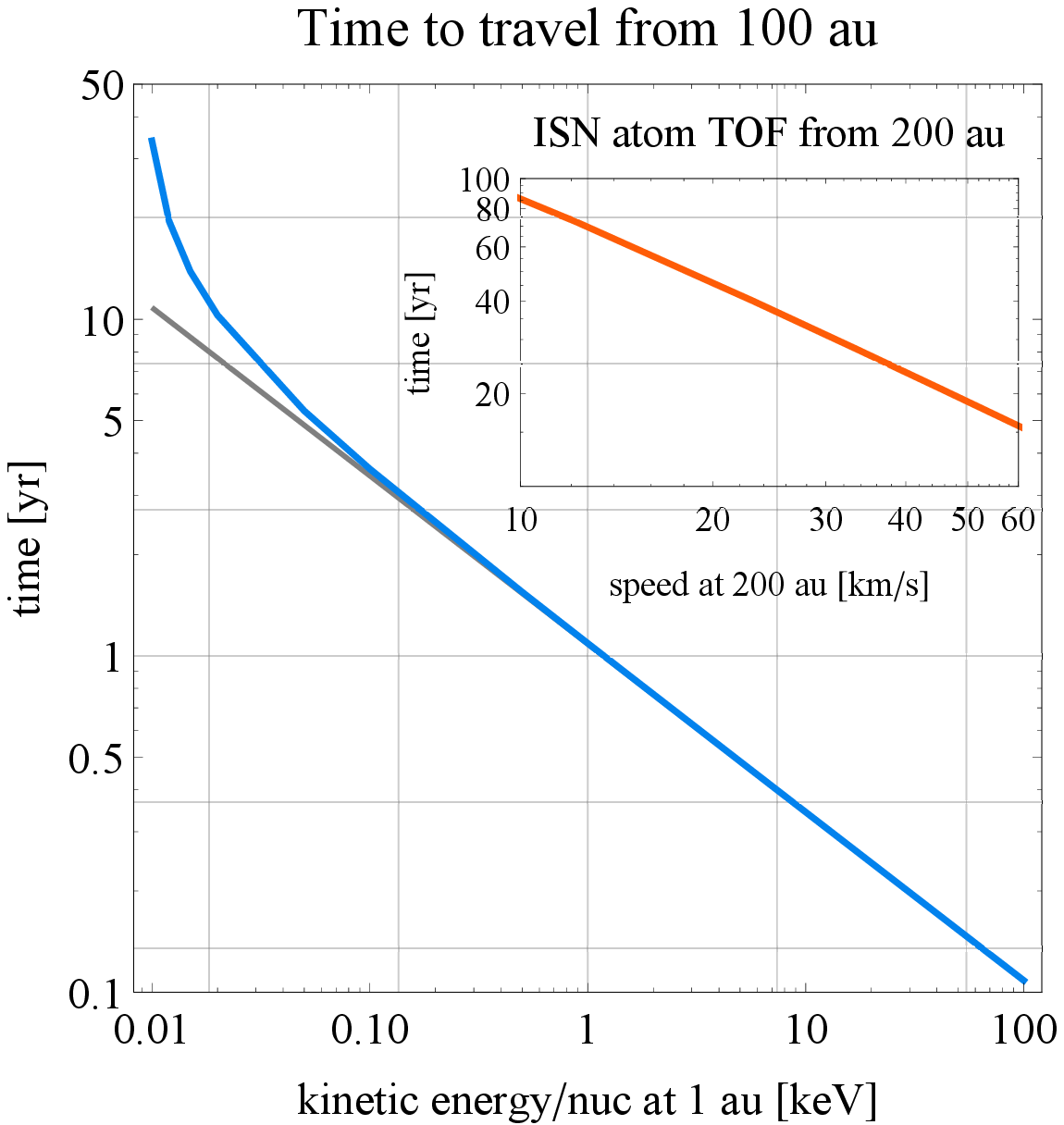} 
\caption{Speed at 100 au (upper panel) and time of flight at a Keplerian trajectory from 100 au to a detector at 1 au (lower panel) for atoms with kinetic energies/nucleon at 1 au ranging from those typical for ISN atoms observed by IBEX-Lo \citep{fuselier_etal:09b} up to those characteristic for the most energetic atoms sampled by the SOHO/HSTOF experiment \citep{hovestadt_etal:95a, hilchenbach_etal:98a}. Effects of acceleration are included. The gray line in the lower panel shows the time of flight for a constant-speed, straight-line motion. Inset in the lower panel shows times of flights (TOF) of He atoms from 200 au for speeds at 200 au from 10 to 60 km~s$^{-1}$}
\label{fig:initSpeed}
\end{figure}

The heliopause in the upwind region is located at $\sim 120$ au \citep{stone_etal:13a, stone_etal:19a}. Secondary ISN atoms originating in the outer heliosheath must cover a distance of $\sim 120 - 250$~au to reach a detector at 1~au. On the other hand, ENAs originating in the inner heliosheath must cover a distance shorter by half, and they travel considerably faster, which suggests that the epochs from which information is brought by the ISN and ENA populations are very different. 

Travel times vs speeds and energies at 100 au for atoms with energies from 0.01 to 100~keV at 1 au are shown in Figure~\ref{fig:initSpeed}. Clearly, for ENAs the approximation of a straight-line, constant-speed motion, frequently used to estimate travel times and distance to the ENA origin sites  in the inner \citep[e.g.,][]{reisenfeld_etal:16a, reisenfeld_etal:19a, mccomas_etal:19a} and outer heliosheath \citep[e.g.,][]{schwadron_mccomas:19a} is fully adequate, and since travel times are relatively short (less/much less than the solar cycle length), it is feasible to connect variations in the solar wind, directly measured at 1 au, with variations in the heliospheric ENA flux. ISN atoms, however, travel much slower ($\sim 25$~ \kms), and the secondary atoms even slower ($\sim 12 - 15$~ \kms). Therefore, it can be expected that their times of flight are much longer and significantly different for the primary and secondary atoms. Because of the acceleration and bending by solar gravity inside the heliosphere, these times of flight cannot be assessed straightforwardly.

ISN atoms observed at 1 au are subjected to time-varying ionization losses \citep{rucinski_bzowski:95b, rucinski_etal:03}, and H atoms additionally to variations in radiation pressure. However, these variations are induced locally from the view point of the global heliosphere,  in a time frame less than a year prior to the observation time \citep{bzowski_etal:02}, which is well inside the time interval covered by available solar wind and solar EUV observations. 

In this paper, we focus on travel times of the primary and secondary ISN atoms of He and H sampled by IBEX and, in the future, by IMAP. We seek to identify the epoch when the presently observed atoms had been produced from the plasma in the outer heliosheath and what is the travel time spread in atom samples observed in different locations in the Earth orbit from different directions in the sky. Answering this question permits to assess a time interval needed for observations of ISN atoms permitting to discover hypothetical spatial variations in the distribution of ISN gas in the local interstellar medium, that would show up as a temporal variation in the gas inflow parameters. It also reveals the epoch that must be covered by time-dependent global models of the heliosphere to make them applicable as a basis for interpretation of ISN atoms observed at 1 au using any observation techniques, be it direct sampling, heliospheric glow, or pickup ion measurements.  

We begin with a brief presentation the method we used for assessing the time of flights, atom arrival dates, and their spread. Subsequently, we discuss the times of flight of the primary and secondary populations of helium atoms observed by IBEX and IMAP. Then, we discuss differences in times of flight between helium and hydrogen atoms and offer predictions, when the ISN atoms that penetrate inside the heliopause at the time when Voyager was exiting it will be visible to a detector at 1 au. We finish with a summary and conclusions.

\section{Calculation of times of flight}
\label{sec:TOFcalculation}
\begin{figure}
\centering
\includegraphics[width=0.5\textwidth]{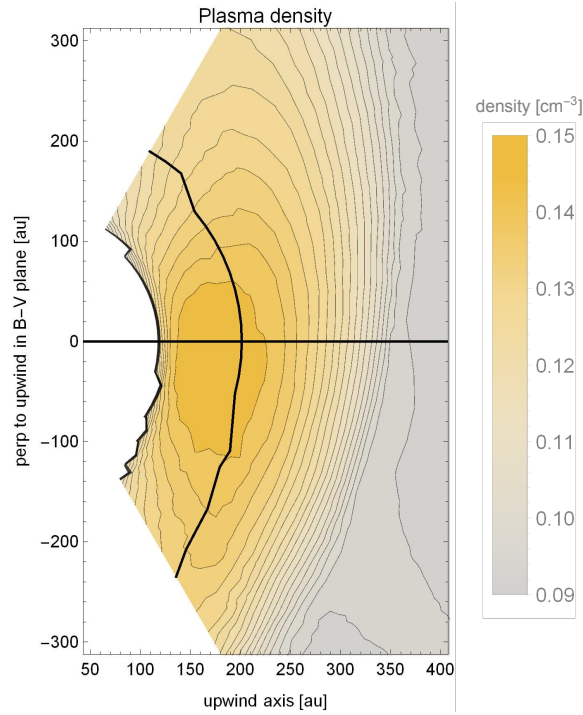}
\caption{Plasma density outside the heliopause in a plane defined by the vectors of Sun's motion and magnetic field in the LISM, determined in \citep{kubiak_etal:16a}, shown to illustrate the region where the production of secondary atoms is the most intense. The thick black lines mark the heliopause at $r\HP$ and the boundary at $r_B = 1.75 \times r\HP$. }
\label{fig:plasmaAcrossDens}
\end{figure}

\noindent
The times of flight of the atoms are calculated using the numerical version of the Warsaw Test Particle Model (nWTPM) simulation code \citep{sokol_etal:15b} with secondary atom synthesis method \citep{bzowski_etal:17a, bzowski_etal:19a}, appropriately adapted for our purposes. We calculate mean times of flight of ISN atoms for specific locations of observations, viewing directions, and instrument fields of view defined by IBEX/IMAP collimator transmission functions. For a given observation date $t\obs$, location $\myvec{r}\obs$, detector velocity relative to the Sun $\myvec{v}\obs$, viewing direction $\myvec{\alpha}$, and an ISN model parameter set $\myvec{\pi}$, the mean time of flight $T_{\text{TOF}}$ from a selected geometric location in space $r_B$ to the detector is calculated as:
\begin{equation}
T_{\text{TOF}}\left( \myvec{r}\obs, \myvec{v}\obs,\myvec{\alpha}, \myvec{\pi}\right) = t\obs - \tau\left(t\obs,\myvec{r}\obs, \myvec{v}\obs,\myvec{\alpha}, \myvec{\pi} \right) \pm \Delta \tau,
\label{eq:TOFDefinition}
\end{equation}
where $\tau$ is the weighted average date of crossing by atoms the interaction region boundary, weighted by observed flux:
\begin{equation}
\tau\left(t\obs,\myvec{r}\obs, \myvec{v}\obs, \myvec{\alpha}, \myvec{\pi} \right)  = T\left(t\obs,\myvec{r}\obs, \myvec{v}\obs, \myvec{\alpha}, \myvec{\pi} \right)/\Phi\left( t\obs,\myvec{r}\obs, \myvec{v}\obs, \myvec{\alpha}, \myvec{\pi} \right)
\label{eq:tauDefinition}
\end{equation}
with
\begin{equation}
T\left(t\obs,\myvec{r}\obs, \myvec{v}\obs,\myvec{\alpha}, \myvec{\pi} \right) = 
\int \limits_{\Delta \alpha}^{} C(\myvec{\alpha})\,d \myvec{\alpha} \int \limits_{u_{\text{min}}}^{u_{\text{max}}}
t_B\,u\,\omega\left(\myvec{v}, \myvec{r}\obs, \myvec{v}\obs, \myvec{\pi} \right)\, u^2 du,
\label{eq:TDefinition}
\end{equation}
\begin{equation}
T_2\left(t\obs,\myvec{r}\obs, \myvec{v}\obs,\myvec{\alpha}, \myvec{\pi} \right) = 
\int \limits_{\Delta \alpha}^{} C(\myvec{\alpha})\,d \myvec{\alpha} \int \limits_{u_{\text{min}}}^{u_{\text{max}}}
t_B^2\,u\,\omega\left(\myvec{v}, \myvec{r}\obs, \myvec{v}\obs, \myvec{\pi} \right)\, u^2 du,
\label{eq:T2Definition}
\end{equation}
\begin{equation}
\Phi\left(t\obs,\myvec{r}\obs, \myvec{v}\obs,\myvec{\alpha}, \myvec{\pi} \right) = 
\int \limits_{\Delta \alpha}^{} C(\myvec{\alpha})\, d \myvec{\alpha} \int \limits_{u_{\text{min}}}^{u_{\text{max}}}
\,u\,\omega\left(\myvec{v}, \myvec{r}\obs, \myvec{v}\obs, \myvec{\pi} \right)\, u^2 du.
\label{eq:PhiDefinition}
\end{equation}
$T$ is the mean calendar date of the boundary crossing, $T_2$ is the mean square of this date, both multiplied by the atom flux at the detector, $\Phi$ is the atom flux at the detector. Note that all these quantities are filtered (weighted) by the collimator function. Similarly, the mean speed $u_B$ at the boundary $r_B$ of the region of interest in the OHS by atoms that reach the detector are obtained as:
\begin{equation}
u_B\left(t\obs,\myvec{r}\obs, \myvec{v}\obs,\myvec{\alpha}, \myvec{\pi} \right) = 
\frac{ \int \limits_{\Delta \alpha}^{} C(\myvec{\alpha})\,d \myvec{\alpha} \int \limits_{u_{\text{min}}}^{u_{\text{max}}}
v_B\,u\,\omega\left(\myvec{v}, \myvec{r}\obs, \myvec{v}\obs, \myvec{\pi} \right)\, u^2 du}
{\Phi\left(t\obs,\myvec{r}\obs, \myvec{v}\obs,\myvec{\alpha}, \myvec{\pi} \right)},
\label{eq:UDefinition}
\end{equation}
where $v_B$ is the speed (i.e., the magnitude of velocity) of individual atoms at crossing the boundary of the region of interest. The boundary $r_B$ is marked with the black contour in Figure~\ref{fig:plasmaAcrossDens}. Note that the $r_B$ distance may be different for different atom trajectories in the calculation.

The integration in the formulae above is carried out in the spacecraft-inertial reference frame, i.e., the motion of the spacecraft relative to the Earth and of the Earth relative to the Sun are taken into account. In this frame, the speed of an atom relative to the spacecraft is $u$. Formally, $u_{\text{min}}=0$ but some of the trajectories would not be able to reach outside the heliopause -- in this case, $\omega$ is set to 0. The lower boundary calculation implemented in the code are based on the formalism presented in Section 2.3.1 in \citet{sokol_etal:15b}. 

The upper boundary is formally $u_{\text{max}} = \infty$, in the code it calculated using the formalism presented in \citet{sokol_etal:15b} to satisfy the condition that the integration in the co-moving frame of the unperturbed ISN gas extends to 4.5 thermal speeds of the gas. In the solar inertial frame, the velocity of an atom hitting the detector is given by $\myvec{v}$, which is a function of $u$ and $\myvec{\alpha}$. The quantity $t_B$ is calendar date of crossing by an individual atom of a certain boundary distance $r_B$ during its travel from the LISM to the detector, $C(\myvec{\alpha})$ is the collimator transmission function, $\omega$ the statistical weight of a given atom, and $\Delta \alpha$ is the solid angle of the instrument field of view. 

We solve the Boltzmann equation using the method of characteristics. Unlike in the Monte Carlo method, our trajectories do not mimic individual atoms. Instead, when we select a given trajectory for calculation, we subsequently calculate the probability of existence at the detector of an atom with the given kinematic parameters against all production and loss processes underway, and given the probability of existence of this atom in the unperturbed LISM. The latter is obtained from the assumed distribution function of neutral gas in the unperturbed LISM. So all trajectories are tracked all the way from the detector back to the LISM and then from the LISM to the detector. The statistical weight $\omega$ is obtained from solution of production and loss balance equation along these trajectories, calculated for each of them separately, using a method presented by \citet{bzowski_etal:17a, bzowski_etal:19a}, and \citet{kubiak_etal:19a}. This calculation goes as follows:
\begin{itemize}
\item in the spacecraft-inertial system, select the atom speed $u$ and the impact direction $\myvec{\alpha}$,
\item calculate the Cartesian velocity vector and transform it to the solar-inertial frame by subtracting the velocity vector of the detector $\myvec{v}\obs$, so that the velocity of the atom relative to the Sun at the detector site is $\myvec{v}$,
\item calculate the atom trajectory from the detection site to the boundary of the calculation region at $r_0 = 1000$~au from the Sun and register the velocity vector of the atom $\myvec{v}_0(t\obs,\myvec{r}\obs, \myvec{v})$,
\item calculate the initial value $\omega_0$ of the statistical weight $\omega$ assuming that the distribution function at the boundary of the calculation region is given by a homogeneous Maxwell-Boltzmann function $f_M(\myvec{v}_0, \myvec{\pi})$, with the interstellar parameters $\myvec{\pi}$ (inflow velocity vector, temperature, and density) adopted from \citet{bzowski_etal:19a}:
\begin{equation}
\omega_0\left(t\obs, \myvec{r}\obs, \myvec{v}, \myvec{\pi}\right)= f_M\left(\myvec{v}_0, \myvec{\pi}\right),
\label{eq:omega0Definition}
\end{equation}
\item identify the milestones at the trajectory: the heliopause at $r\HP$ and the location within the interaction region at $r_B = 1.75 \times r\HP$; register the time and speed of the atom at those milestone points,
\item solve the production and loss balance equation along the trajectory between the point of entrance to the simulation region and the detector \citep[see Equation~4 in][]{bzowski_etal:19a}
\item insert the solution for $\omega$ to Equations \ref{eq:TDefinition}--\ref{eq:UDefinition}.
\end{itemize}

For He atoms, the production terms are due to charge-exchange injection of neutralized interstellar He$^+$ ions into trajectories leading to the detector \citep[Equations 4, 5 in][]{bzowski_etal:19a}. This process stops at the heliopause. The loss terms include ionization by solar EUV radiation and by charge-exchange collisions between neutral He atoms at a trajectory leading to the detector and the ambient He$^+$ ions \citep[Equations 8, 9 in][]{bzowski_etal:19a}. Inside the heliopause, the loss terms due to photoionization, charge exchange, and electron-impact ionization are included, with the rates adopted from \citet{sokol_etal:19a}. 

In the simulation, the vector $\myvec{v}_0$ is obtained from solution of equation of motion. For He atoms, $\omega_0$ does not depend on $t\obs$ because the trajectory is given by a purely Keplerian solution, but for H, the total is a sum of the solar gravity force and the force of radiation pressure, which is a function of the time-varying solar Lyman-$\alpha$ spectral flux \citep{IKL:20a}. Consequently, the equation of motion must be solved numerically \citep{tarnopolski_bzowski:09} and the trajectories and times of flight may be different for identical locations in space but different times.  

For ISN H atoms, the production and loss terms outside the heliopause are set to 0 and inside the heliopause, only loss terms are included, calculated based on a model by \citet{sokol_etal:19a}. We assume that the primary and secondary populations exist already at the boundary of the simulations and that neither of them is modified inside the outer heliosheath; this corresponds to the ``two-Maxwellian approximation'' from \citet{kubiak_etal:19a}. 

The integration over the detector field of view is carried out within a solid angle region $\Delta \alpha$ delineated by the collimator boundary, centered at a direction $\myvec{\alpha}$, with weighting given by the collimator transmission function defined in Equations 31, 32 in \citet{sokol_etal:15b}. For the location where calendar dates $t_B$ or $t\HP$ are taken, we adopt intersections of individual trajectories leading to the detector either with the heliopause at $r\HP$, or with a surface defined by the surface of the heliopause blown up to a distance $r_B = 1.75 \times r\HP$ (see Figure~\ref{fig:plasmaAcrossDens}). This latter choice was adopted because most of the production of the secondary atoms occurs in this region, as shown by \citet{kubiak_etal:19a}.

Atoms making up an IBEX/IMAP signal within an individual spin angle interval feature a spread in the arrival  direction and in speed. Consequently, the times of flight also feature a spread, which we calculate as the second central moment:
\begin{equation}
 \Delta \tau = \sqrt{T_2\left(t\obs,\myvec{r}\obs, \myvec{\alpha}, \myvec{\pi} \right)/\Phi\obs\left(t\obs,\myvec{r}\obs, \myvec{\alpha}, \myvec{\pi} \right) - \tau^2\left(t\obs,\myvec{r}\obs, \myvec{\alpha}, \myvec{\pi} \right)}
\label{eq:TOFSpreadDefinition}
\end{equation}

For He, the production and loss terms are obtained based on a plasma flow outside the heliopause taken from the Huntsville model of the heliosphere \citep{pogorelov_etal:09e, heerikhuisen_pogorelov:10a} with the interstellar parameters identical to those used by \citet{kubiak_etal:19a} to study the distribution function of ISN He in the outer heliosheath. It is important to realize that the WTPM synthesis method does not formally differentiate between the primary and secondary atoms.  

The simulations were performed assuming LISM parameters identical to those used by \citet{bzowski_etal:19a} (see their Table~1); in particular, the bulk speed was assumed 25.4~\kms, and the 7500~K, identical for ISN He and for the primary population of ISN H. The inflow direction for both these species was adopted $(255.7\degr, 5.1\degr)$ in the ecliptic J2000 coordinates. The time of flight simulations for ISN He were performed for selected IBEX orbits from ISN observation season 2010 : 54 -- where almost solely the secondary population of ISN He is observed, November 22; 61 -- where a mixture of the primary and secondary is visible, January 14; 64 -- where the primary population dominates and the secondary is only visible at the far wings of the signal, February 6; and for orbit 68 -- where IBEX sampled a different portion of the primary ISN distribution function and the secondary population is still visible in the wings of the signal, March 9. For ISN H, we simulated orbit 68, orbits 69 and 70, where the ISN H signal is the most conspicuous \citep{galli_etal:19a, rahmanifard_etal:19a} and the flux of the primary population is expected to exceed that of the secondary. In the following, we only show the results for orbit 68 and equivalent during later years because for the other ones, the results are almost identical.

\section{Results}
\label{sec:results}

\subsection{ISN He}
\label{sec:resultsHe}
\begin{figure*}
\centering
\includegraphics[width=0.48\textwidth]{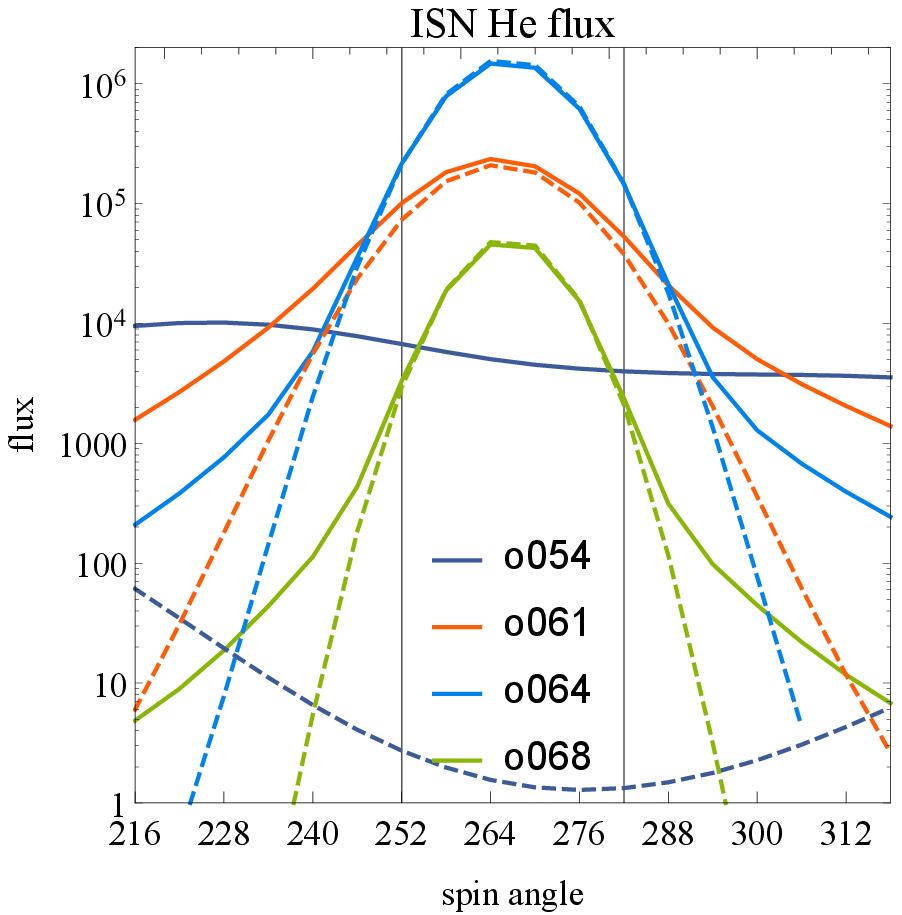}
\includegraphics[width=0.46\textwidth]{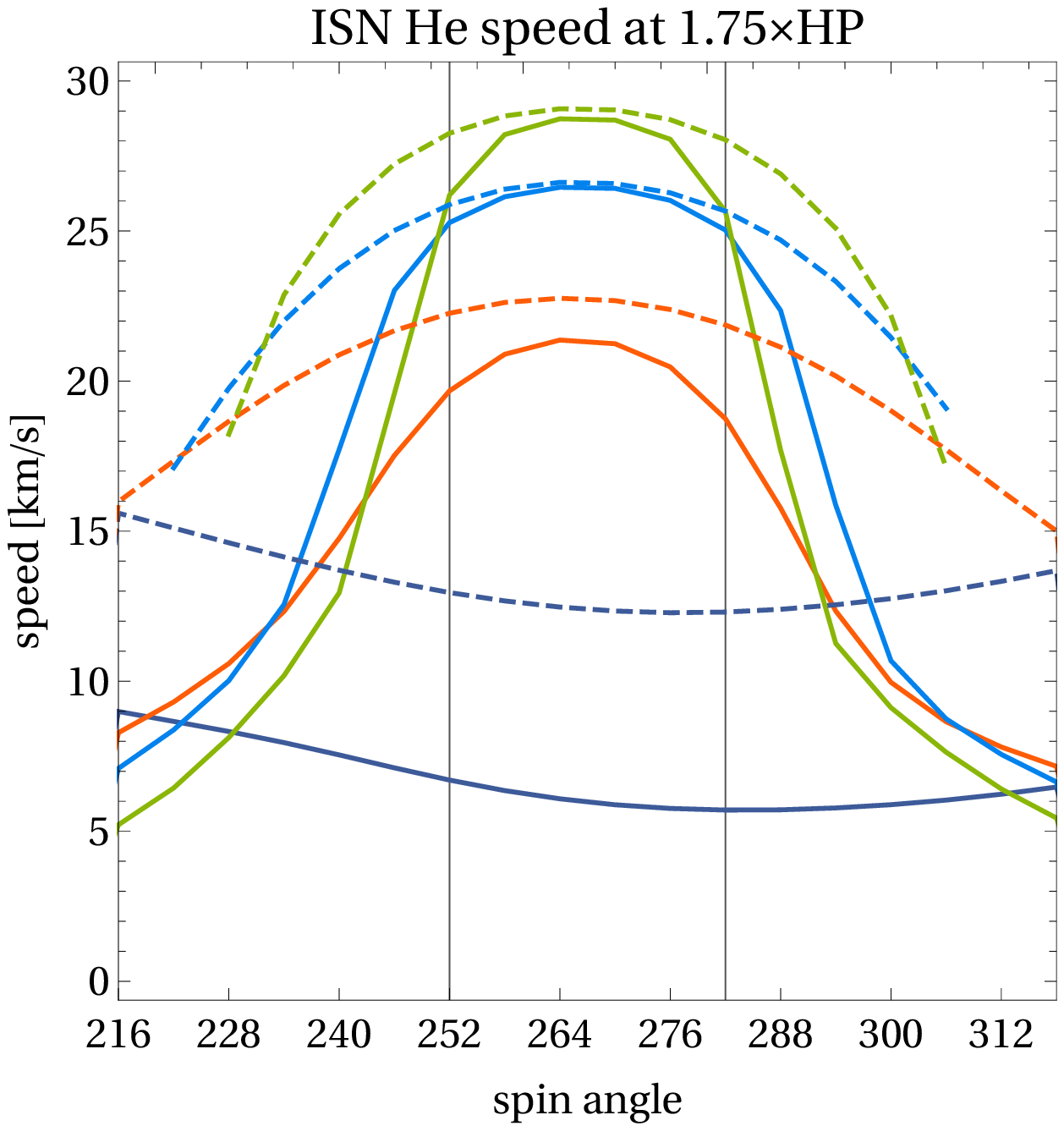}

\includegraphics[width=0.48\textwidth]{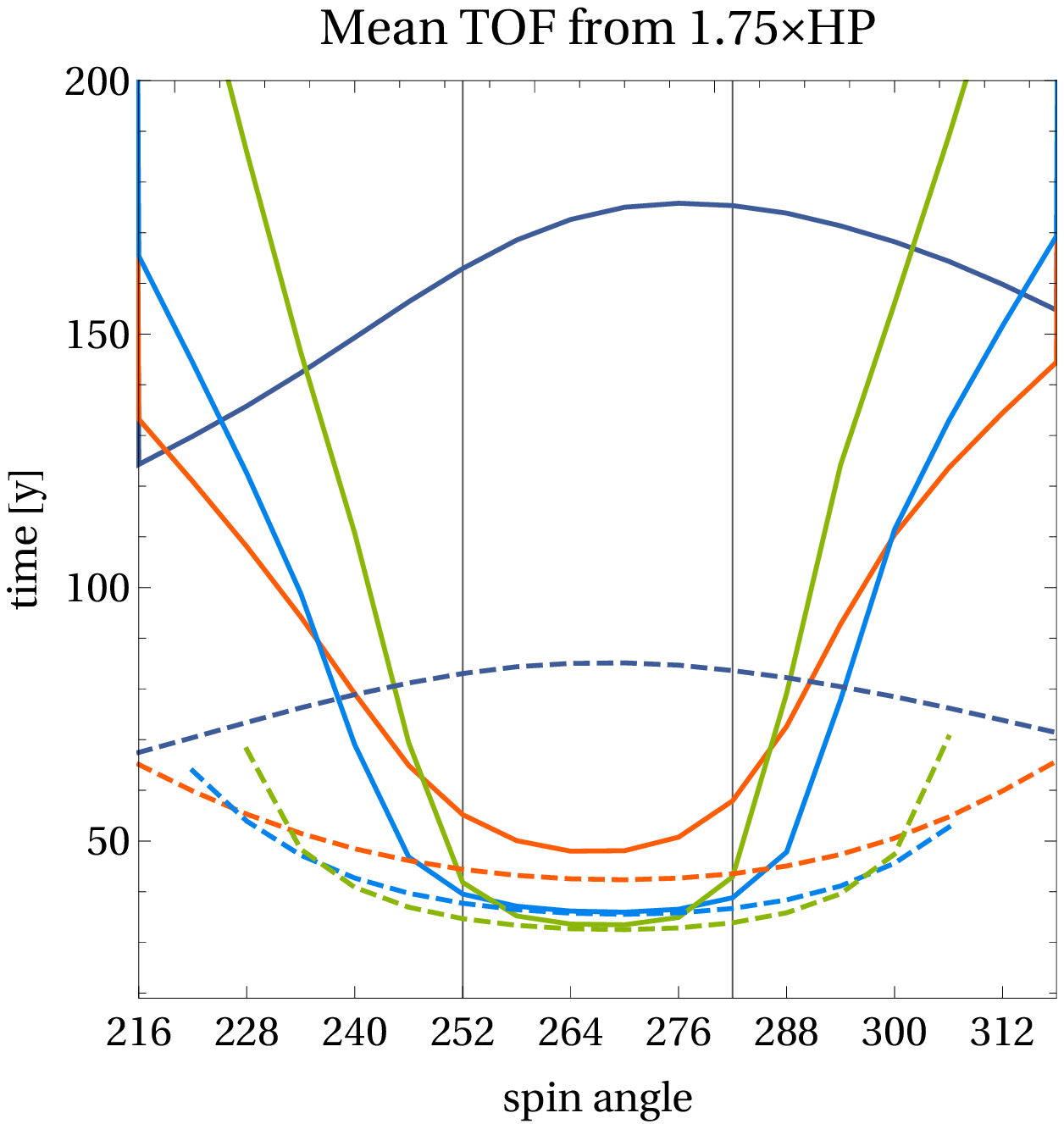}
\includegraphics[width=0.48\textwidth]{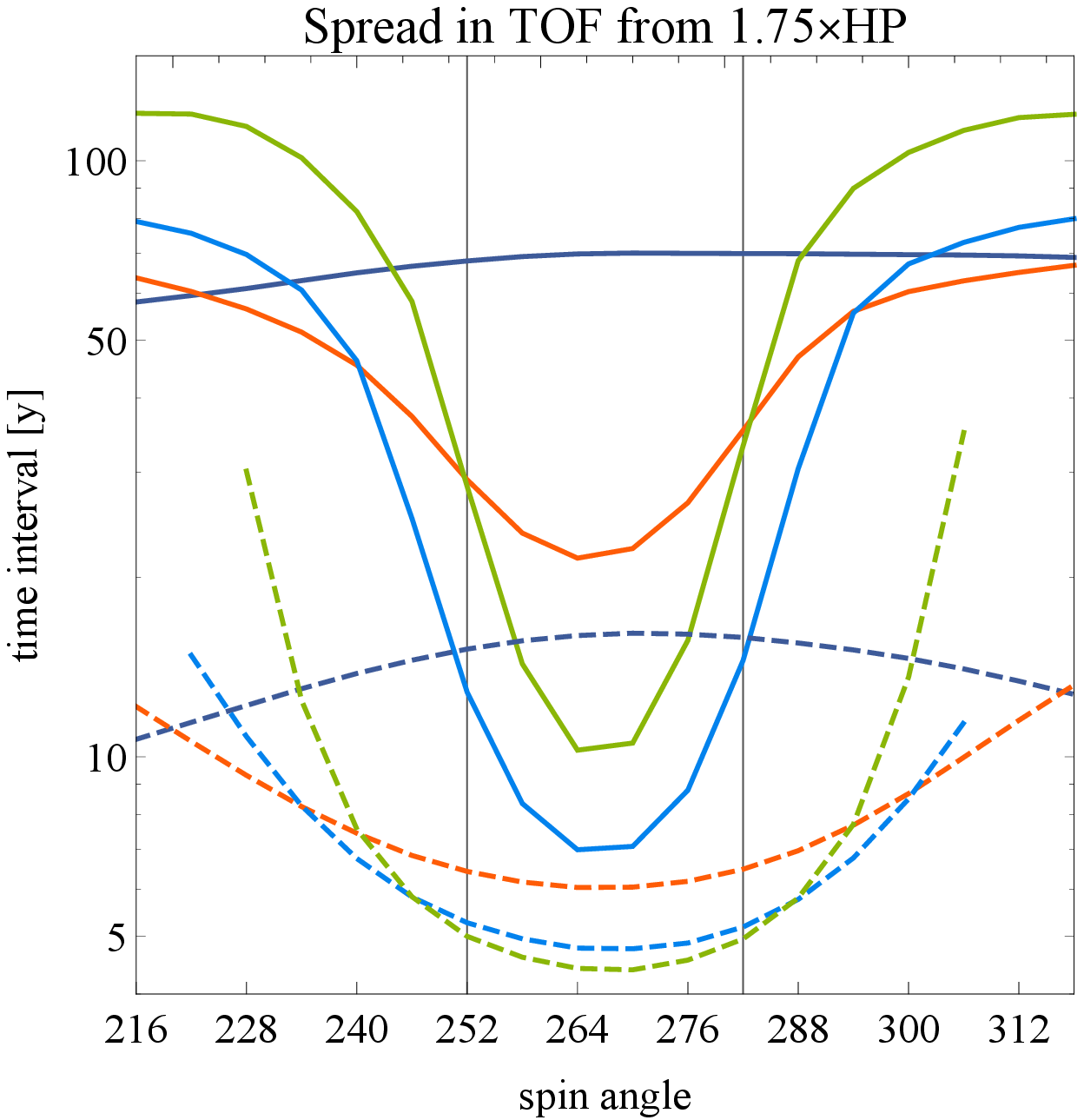}
\caption{Flux of ISN He for four selected IBEX orbits: 54, 61, 64, and 68 (upper left), average speed at $1.75\times r\HP$ (upper right), mean time of flight from $1.75\times r\HP$ to the detector (lower left) and the spread of the mean times of flight (lower right). Solid lines correspond to the model with secondary atom synthesis method, the broken lines to a model where the production and loss terms in the OHS are set to 0. These latter simulations are provided for reference. The two vertical bars represent the spin angle range used by \citet{bzowski_etal:15a} and \citet{swaczyna_etal:18a} to fit the temperature and the flow vector of the primary population of ISN He.}
\label{fig:heIBEX}
\end{figure*}

\begin{figure*}
\centering
\includegraphics[width=0.48\textwidth]{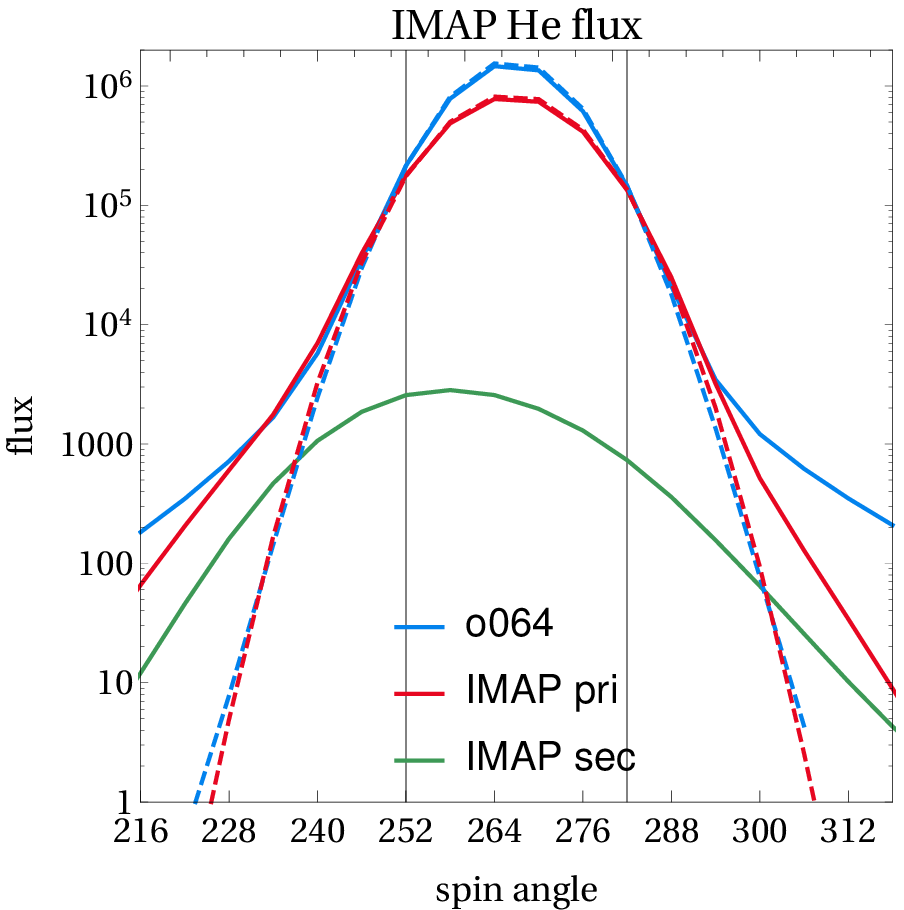}
\includegraphics[width=0.46\textwidth]{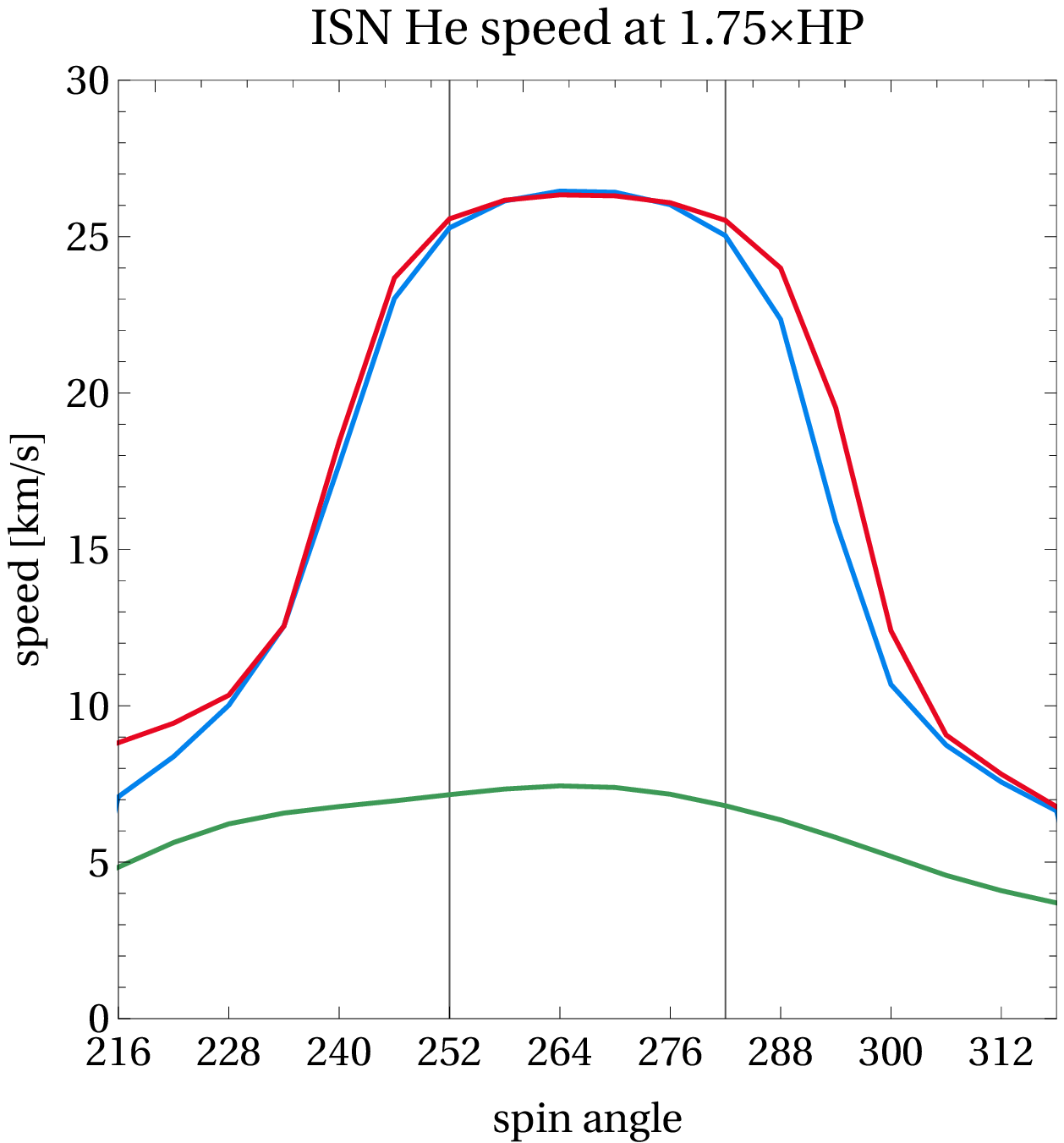}

\includegraphics[width=0.48\textwidth]{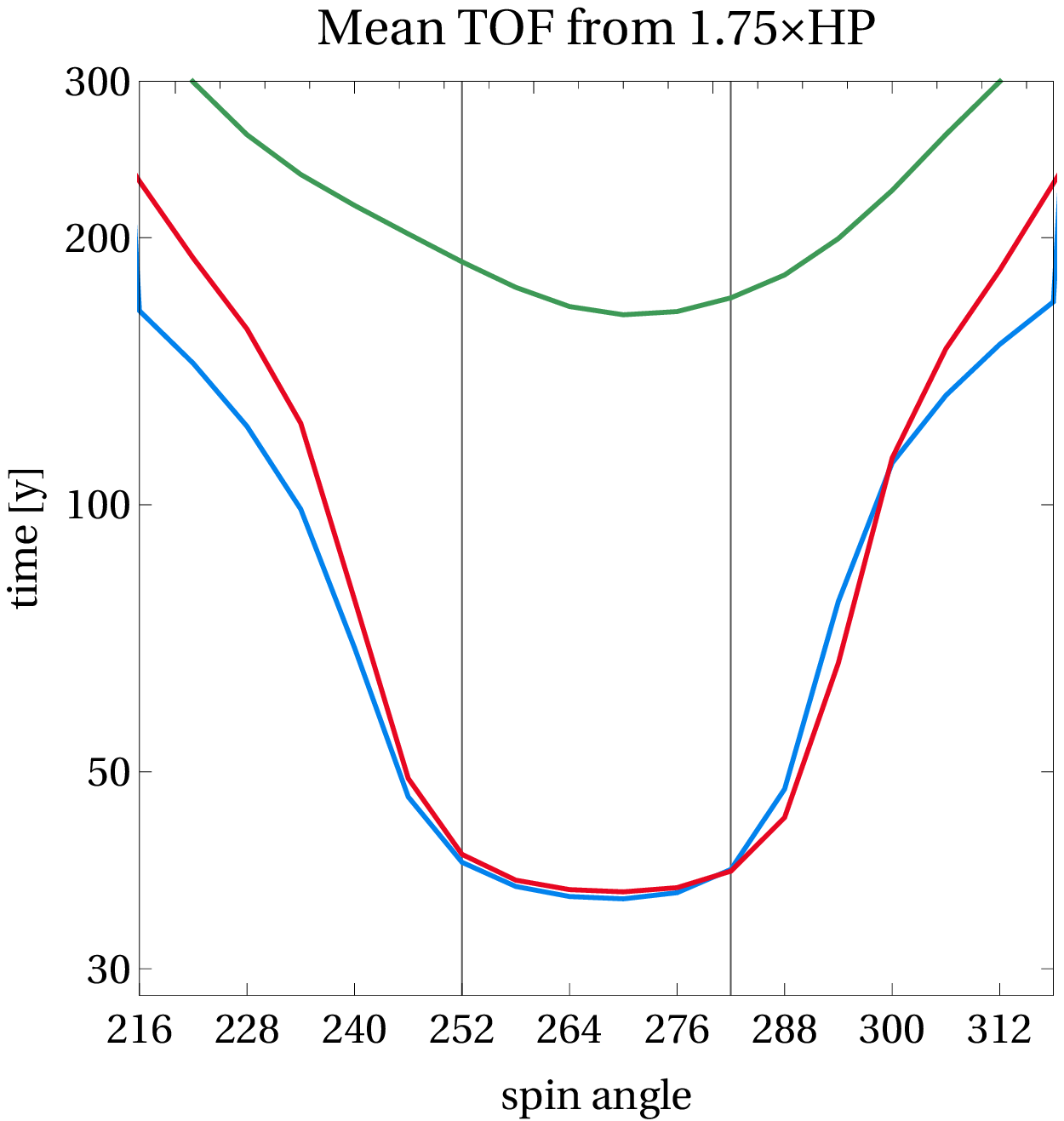}
\includegraphics[width=0.48\textwidth]{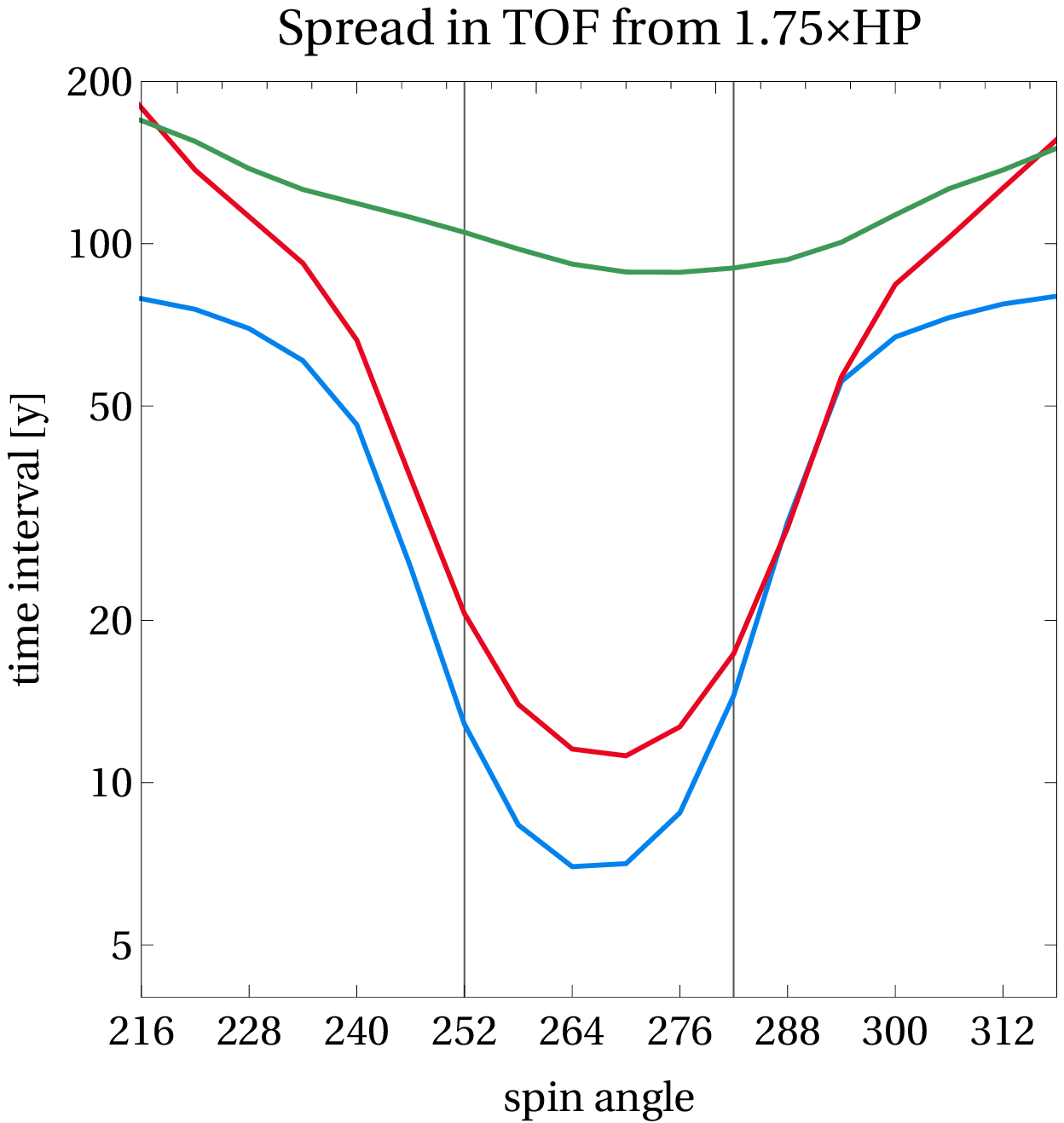}
\caption{Similar to Figure~\ref{fig:heIBEX} but for a geometry of direct-sampling observations enabled in the future by IMAP. IBEX orbit 64 is repeated from the aforementioned figure and compared with two possible observations by IMAP, one optimized for sampling mostly the primary population (IMAP pri; DOY 113, boresight solar elongation 128\degr) of ISN He, and the other for sampling mostly the secondary population of He (IMAP sec; DOY 49, elongation 124\degr). The broken lines in the flux panel (upper left) correspond to the pure ISN He population (with no interaction in the OHS), simulated for IBEX and for the IMAP geometry optimized for the primary population. In comparison with the corresponding solid lines, they illustrate the amount of the secondary atoms, always present in the observed samples. The green broken line is missing because it is below the frame of the figure (i.e., practically no primary population expected in this viewing geometry). The two vertical bars represent the spin angle range used by \citet{bzowski_etal:15a} and \citet{swaczyna_etal:18a} to fit the temperature and the flow vector of the primary population of ISN He. }
\label{fig:heIMAP}
\end{figure*}

\noindent
We provide context for the discussion of times of flight by presenting the simulated flux of ISN He for the four selected orbits in the upper-left panel of Figure~\ref{fig:heIBEX} (solid lines). This flux is obtained using the secondary atom synthesis method. The atoms contributing to the flux observed in individual spin angle bins belong both to the primary and the secondary populations -- this method does not differentiate between the populations, similarly to the measurements. To facilitate assessing the contribution from the secondary population, we also show the flux simulated assuming no interaction between ISN He atoms and interstellar He$^+$ ions in the OHS (unperturbed primary population -- broken lines). The secondary population can be approximately regarded as the difference between the respective solid and broken lines. The secondary atoms clearly dominate in the wings of the flux distribution (the solid lines are high above the broken lines), but also at the cores, where the primary population dominates, a certain amount of the secondary atoms is present for all orbits. In orbit 54, i.e., one of the orbits where \citet{kubiak_etal:14a} discovered the secondary population of ISN He (dubbed the Warm Breeze), practically the entire population observed by IBEX consists of the secondary atoms, as illustrated by the large difference between the solid and the broken dark-blue lines for this orbit in the upper-left panel of Figure~\ref{fig:heIBEX}. 

The atoms illustrated in Figure~\ref{fig:heIBEX} are a subset of all ISN He atoms at the respective locations along the Earth orbit. This is because of the observation conditions: the instrument is traveling with a certain velocity in space, which modifies the population of atoms able to enter the collimator, and additionally, the collimator filters the atoms that have entered the instrument aperture. Details are discussed in \citet{sokol_etal:15b}. Therefore, the times of flight and the other quantities discussed further on are not always characteristic for the entire ISN population at the detector location. They are, however, characteristic for the atoms that bring the information to the detector. Furthermore, the atoms at 1 au are not necessarily the most typical for the atoms present in the OHS in the interaction region. This is illustrated by \citet{kubiak_etal:19a}, where speed-integrated distribution functions for selected locations within the OHS are presented. These authors demonstrate that only a small subset of the atoms with specific parameters of motion are able to reach 1 au at an angle that enable them to enter the aperture of the instrument (cf Figures 3 and 8 from their paper). However, these atoms are the carriers of the information that the detector is gathering; all other atoms are not interesting for the topic of this paper because they do not contribute data for our understanding of the process operating in the OHS.

In the lower-left panel of Figure~\ref{fig:heIBEX} we show that the times of flight of IBEX He atoms from the interaction region in the OHS ($T_{\text{TOF}}$ defined in Equation~ \ref{eq:TOFDefinition}) are very diversified: from about 30 years for the primary atoms, up to 200 years for the secondary atoms. As secondary atoms we include those observed in the portions of Earth's orbit where IBEX samples almost solely the secondary population, as in orbit 54, and those in the far wings in the other presented orbits. Note that the secondary population is observed in all orbits, including those where the signal from the primary population dominates. The conclusion is that, in the spin angle range corresponding mostly to the primary population (spin angles 250\degr -- 282\degr, grey bars in Figure~ \ref{fig:heIBEX}), IBEX observes atoms that passed the interaction region approximately three solar cycles earlier. In the spin angle range corresponding mostly to the secondary population, the delay is much longer: $\sim 50$ -- 200 years, i.e., $\sim 5$ -- 20 solar cycles. In addition to the long times of flight, the spread of the times of flight ($\Delta \tau$ in Equation~\ref{eq:TOFSpreadDefinition}) is also large: from a half of the solar cycle period for the primary atoms to 50 -- 100 years (5 -- 10 solar cycle periods) for the secondary atoms.

Inspection of the upper-right panel in Figure~\ref{fig:heIBEX} reveals that mean speeds at $r_B = 1.75 \times r\HP$ of the atoms contributing to the IBEX signal ($u_B$ defined in Equation~\ref{eq:UDefinition}) differ for the two cases considered (i.e., with the OHS interaction included or switched off), even for the spin angle bins containing mostly the primary atoms. This is evidence that in direct-sampling observations at 1 au, there are no ``pure primary population'' observations: a contribution from the secondary and/or effects of filtration of the primary are always present. If anything, then one can select orbits and spin angle ranges where almost solely the secondary population is present \citep[see][]{kubiak_etal:16a}. 

The magnitude of the flux and its variation with IBEX spin angle evolve during the observation season. An important factor shaping this evolution is the geometry of IBEX observations, with the spacecraft spin axis maintained within a few degrees from the Sun and viewing direction perpendicular to the spacecraft--Sun line \citep{mccomas_etal:09a}. IBEX is able to see atoms that belong to the core of the distribution  within the OHS only during  one orbit each year (in our selection, orbit 64). However, the conclusion that times of flight for the primary atoms are on the order of three solar cycles and for the secondary atoms an order of magnitude longer holds regardless of the viewing geometry. We verified this by simulating a different viewing geometry, namely that planned for a future space experiment IMAP-Lo \citep{mccomas_etal:18a, sokol_etal:19c}. In this geometry, the viewing direction of the detector is no longer fixed perpendicular to the spacecraft rotation axis, and consequently, it will be possible to maintain the peak of the ISN flux in the instrument field of view almost throughout the year. \citet{sokol_etal:19c} identified advantageous viewing geometries. We selected one of them, distant from the vantage points of IBEX, and calculated the times of flight and their spread, as well as the speeds of the observed atoms in the OHS. In addition to the viewing geometry optimized for observing the primary atoms, we performed similar calculations for the viewing geometry optimized for observing almost solely the secondary atoms. Results are presented in Figure~\ref{fig:heIMAP}. 

Clearly, when the detector is looking into the peak of the flux of incoming atoms, times of flight of the observed primary atoms are little dependent on the location of the detector along the Earth orbit. These shown in Figure~\ref{fig:heIMAP} are similar to those shown in Figure~\ref{fig:heIBEX}. For the secondary atoms, times of flights do depend on the location of the detector and the viewing geometry; in the geometry selected in Figure~\ref{fig:heIMAP}, the times of flight of the secondary atoms are even longer than for IBEX orbit 54 and vary from 200 to 300 years, with a spread from 100 to 200 years.

\subsection{ISN H}
\label{sec:resultsH}
\begin{figure}
\centering
\includegraphics[width=0.48\textwidth]{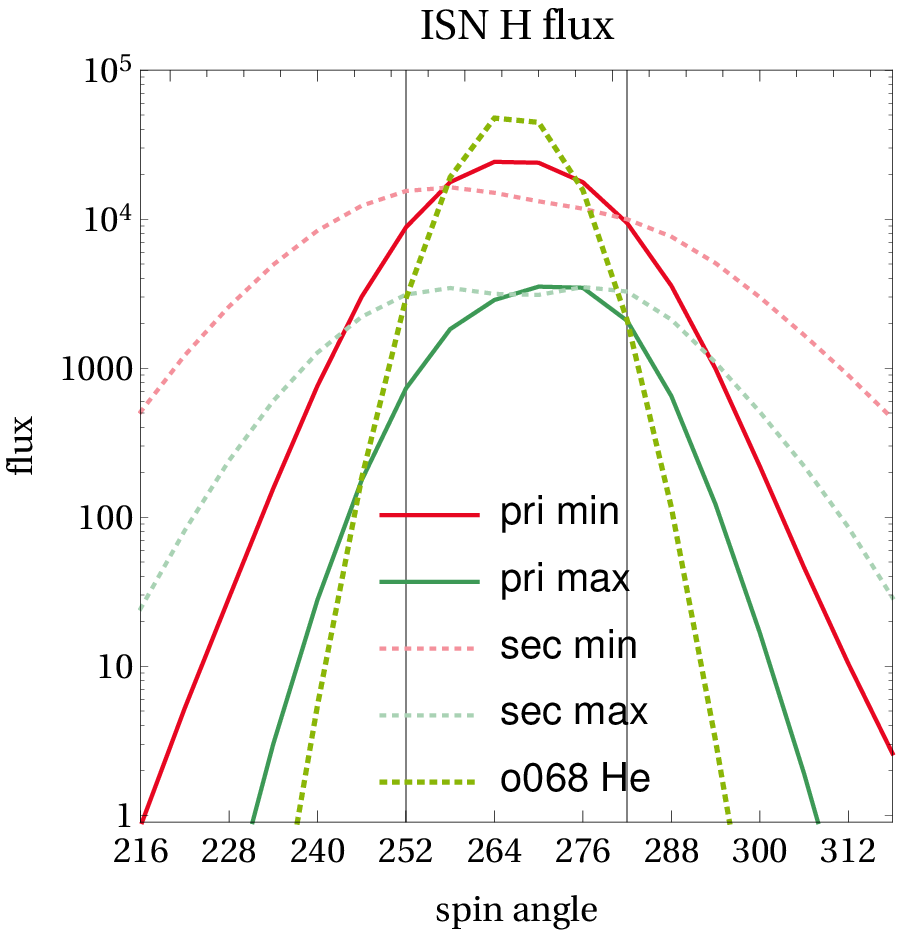}
\includegraphics[width=0.48\textwidth]{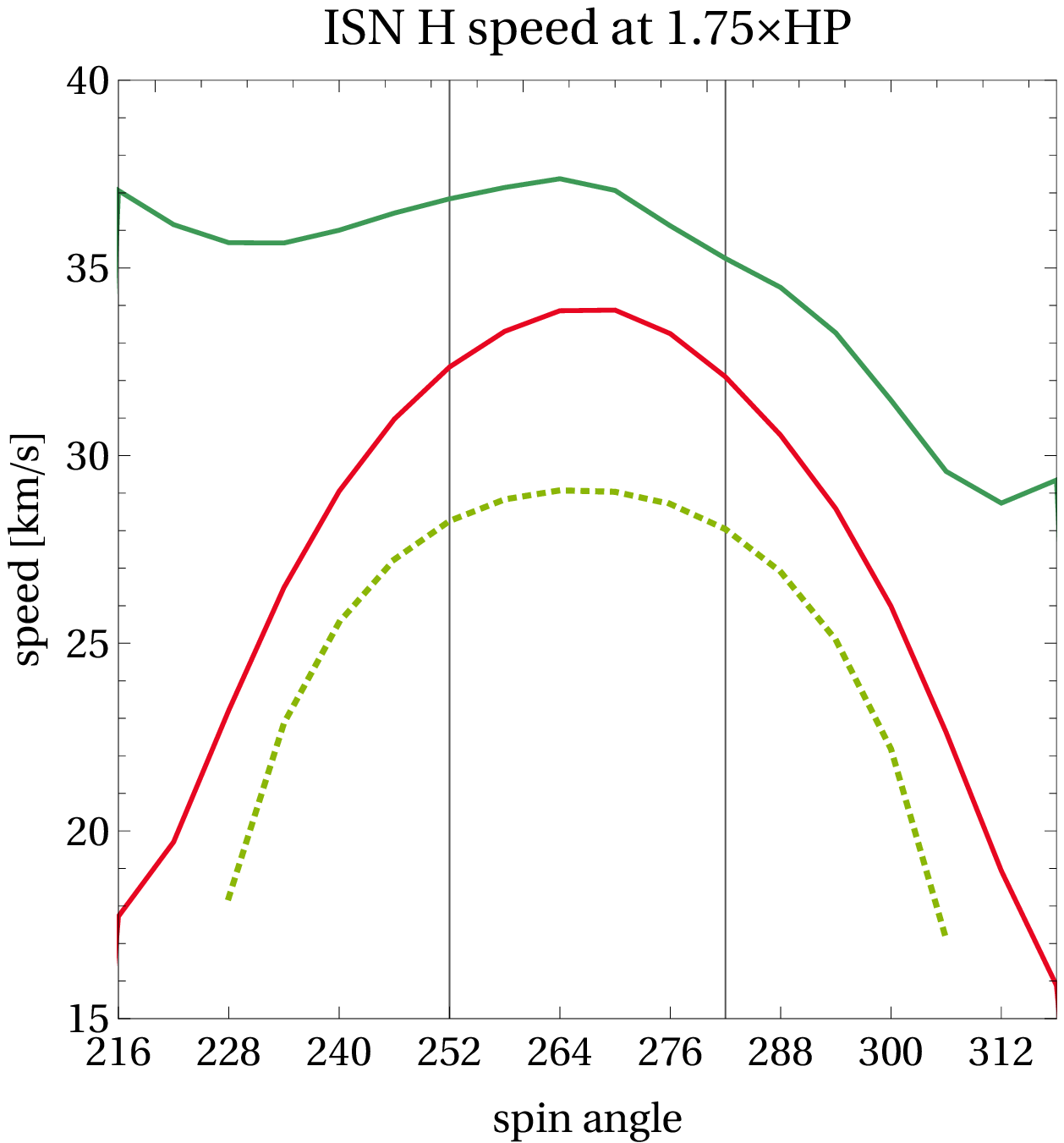}

\includegraphics[width=0.48\textwidth]{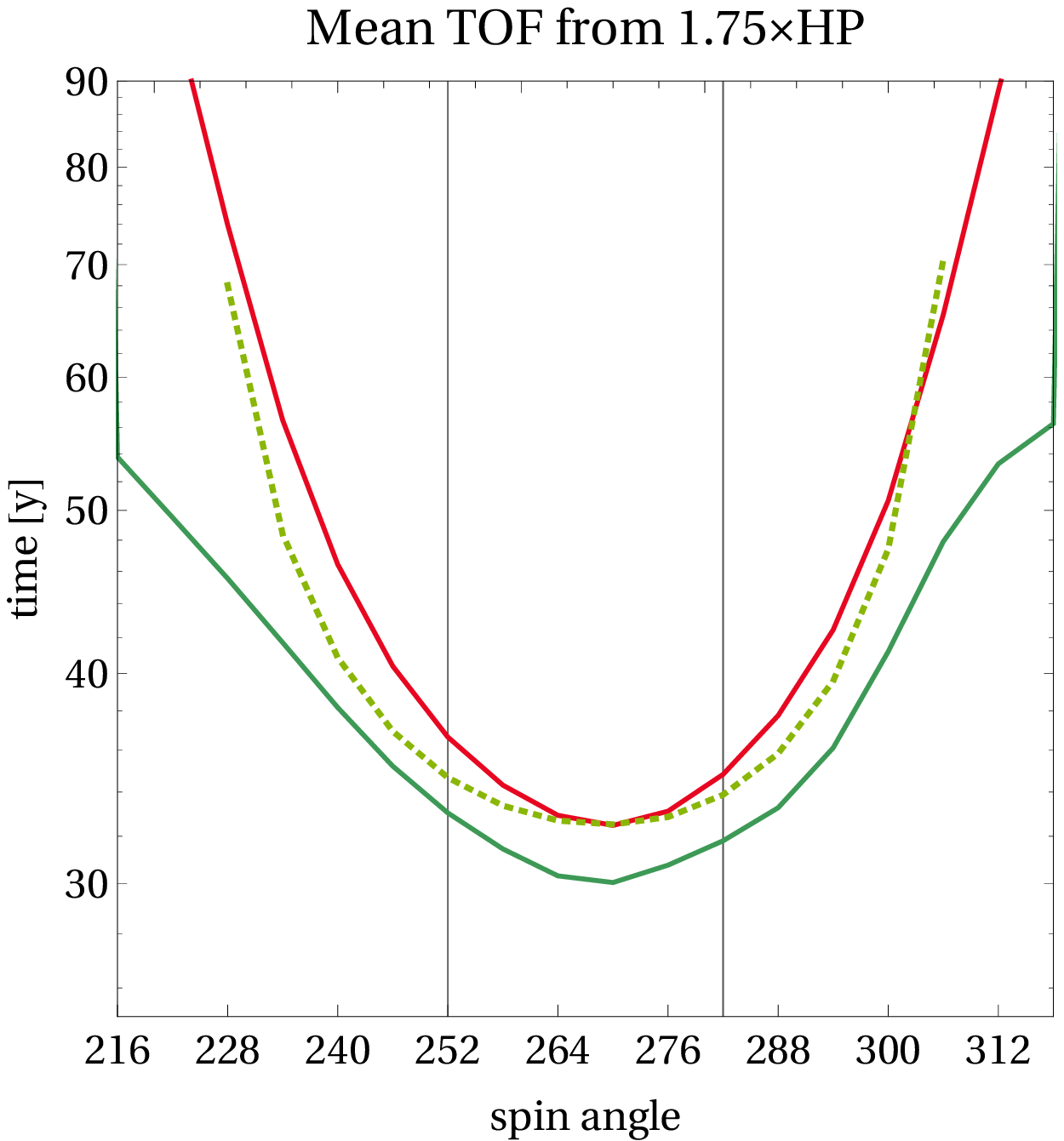}
\includegraphics[width=0.48\textwidth]{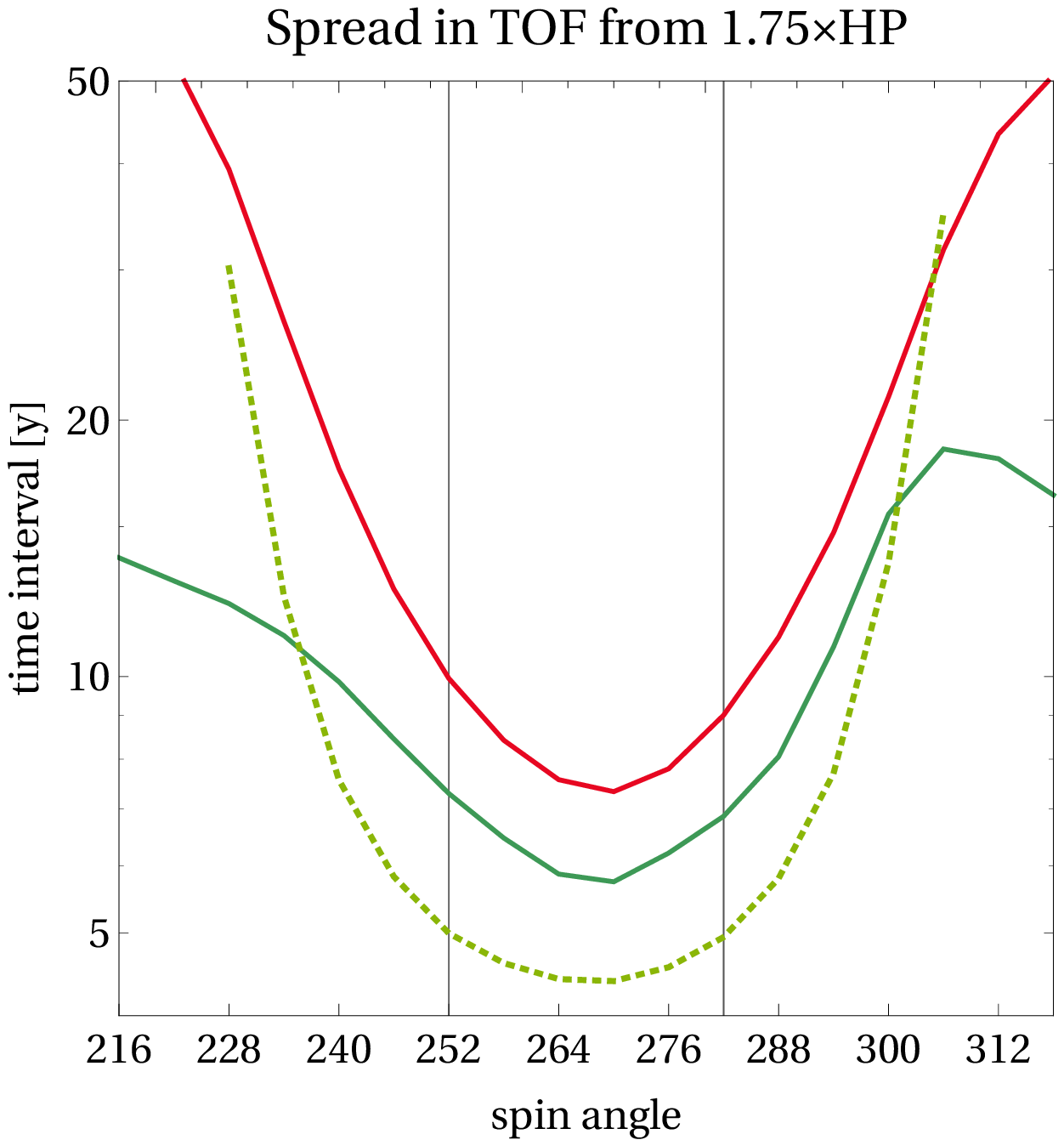}
\caption{Flux of ISN H (primary and secondary populations, solar minimum  and maximum) at IBEX, compared with the flux of ISN He (upper left panel), shown to demonstrate the very limited range of spin angles where the primary population exceeds the secondary. Viewing geometry characteristic for IBEX orbit 68 and its equivalent for the observation season 2015. Times of flight from the interaction region at $r_B = 1.75\times r\HP$ (lower-left panel), atom speed at $r_B$ (upper right), and the spread in the times of flight for the primary population of ISN H during solar minimum nad maximum, all compared with the corresponding quantities for ISN He. }
\label{fig:hIBEX} 
\end{figure}
\begin{figure*}
\centering
\includegraphics[width=\textwidth]{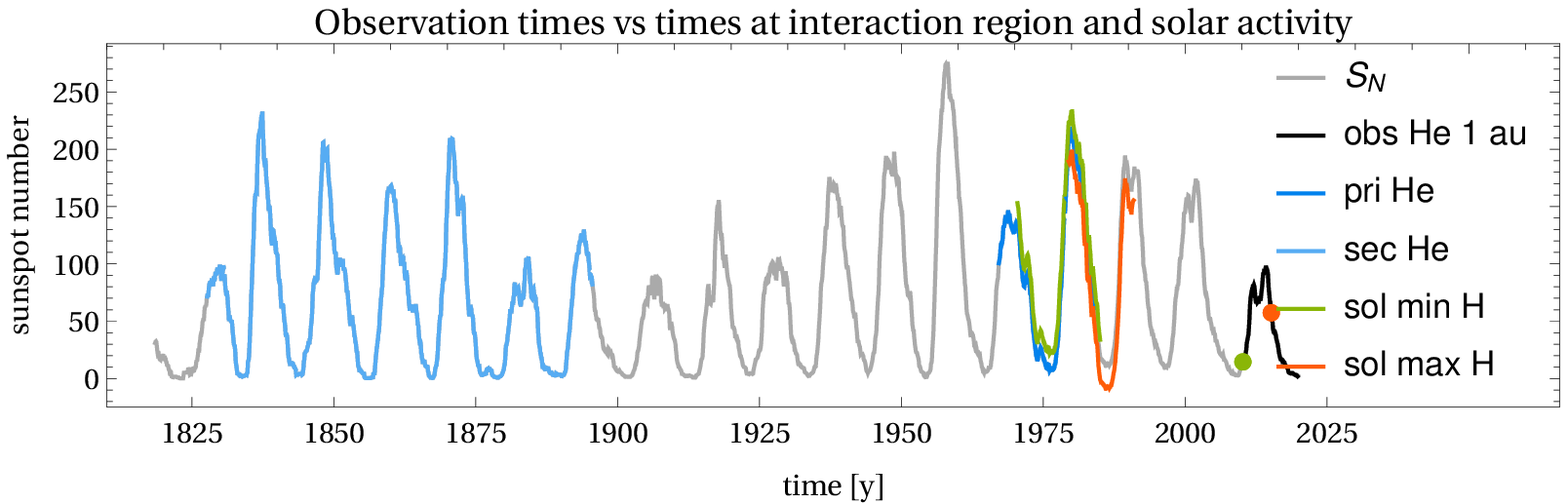}
\caption{{\emph{fig:sunspots}} Compilation of approximate differences between IBEX observation times of ISN atoms and the times of the atoms penetration of the interaction region in the OHS, put in the context with the history of the solar activity. The solar activity level is illustrated by a smoothed revised Brussels sunspot number $S_N$ \citep[gray line,][]{clette_etal:16a, clette_lefevre:16a}. The observation interval of ISN He by IBEX (2009--2020) is shown with the black line, and the two selected observations of the primary ISN H with the green point for the solar minimum (2010), and the orange point for the solar maximum conditions (2015). The time interval when the primary population of ISN He observed in 2009---2020 was penetrating the interaction region in the OHS is marked in blue, and the penetration interval for the secondary He in pale blue. The times of penetration of the interaction region by the primary population of ISN H observed in 2010 are marked in green, and those observed in 2015 in orange.}
\label{fig:sunspots}
\end{figure*}

We are able to provide only a limited insight into ISN H because WTPM with synthesis method cannot be reliably applied in a situation where the secondary population is larger than the primary, as seems to be the case with ISN H \citep[see, e.g.,][]{izmodenov_etal:09a}. Therefore, we limit the discussion to the orbits where the signal from the primary population of ISN H is expected to be larger than that from the secondary, i.e., 68, 69, and 70 and their equivalents for the solar maximum conditions (for the ISN observation season 2015). For the primary H at the entrance to the simulation region, we adopt identical temperature and flow vector as those for ISN He. In the following, we only show the times and speeds for the viewing geometry for orbit 68 because, as we found, these quantities for orbits 69 and 70 are almost identical. 

Additional factors, important for ISN H but absent for ISN He, are on one hand the solar Lyman-$\alpha$ radiation pressure, which modifies the trajectories of individual atoms, and on the other hand ionization losses in the inner heliosphere, much larger than those for He \citep{rucinski_etal:96a}. Radiation pressure compensates for solar gravity during low solar activity (2010), and overcompensates the gravity during high solar activity (2015), which makes the effective force repulsive. Intense ionization masquerades for dynamical effects due to selective elimination of slower atoms, leaving only the faster ones in the sample. This results in speeding up of the entire population at 1 au and a change in the mean flow direction \citep{bzowski_etal:97}. By contrast, within the interaction region in the OHS, the primary ISN H atoms are expected to flow almost independent on the phase of the solar cycle, similarly to ISN He. 

The speeds of atoms within the interaction region in the OHS are expected to be independent on the solar activity level. The speeds in the interaction region of the collimator-averaged subset of primary ISN He atoms that enter the detector in the spin angle bin 264\degr~(i.e., with the maximum flux) are indeed close to those assumed at the boundary of the simulation region (dotted line in the upper right panel of Figure~\ref{fig:hIBEX}). This is not the case for H atoms. The H atoms observed during solar minimum had their speeds in the interaction region larger by $\sim 7$~\kms{} than the assumed bulk velocity, and those observed during high solar activity still larger. This latter effect seems counter-intuitive, because the high radiation pressure could be expected to slow down the entire population at 1 au. This is not the case, however, because due to the high radiation pressure only a subset of atoms can reach 1 au, namely those with a sufficiently high speed in the interaction region. This speed must be high enough to prevent the increased radiation pressure from repelling them altogether. Since these latter atoms travel faster most of the way, their time of flight from the interaction region is shorter than that of He atoms and of the H atoms observed during solar minimum, and their averaged speed is larger.

ISN H atoms observed during solar minimum have a little longer times of flight but larger speeds in the source regions than ISN He atoms because of the radiation pressure action discussed above. Additionally, during all epochs, a part of this effect is induced deep inside  the heliosphere due to selective ionization: the atoms that traveled slower had been eliminated more readily than the faster ones, but while only the faster ones survived, the mean time of flight of the observed sample is a a little longer. 

In all,  however, the delay for the primary population of ISN H is similar to that for ISN He and close to at least three--four solar cycle periods, with variations between low and high solar activity on the order of half solar cycle length.

\subsection{Epochs of penetration of the interaction region}
\label{sec:penetrationEpochs}
\noindent
In this section, we consider the epochs when the atoms presently observed by IBEX traversed the interaction region in the OHS. A simplified compilation of the delay between penetration of the interaction region in the OHS and detection is presented in Figure~\ref{fig:sunspots}, put in context with variations of the solar activity, represented by smoothed modified sunspot number $S_N$ \citep{clette_etal:16a, clette_lefevre:16a}.

For a given detection time $t_d$, we calculate the time span of penetration of the interaction region $(t\start, t\stopp)$ as:
\begin{eqnarray}
t\start & = & t_d - T_{\text{TOF}} - \Delta \tau \nonumber \\
t\stopp  & = & t_d - T_{\text{TOF}} + \Delta \tau.
\label{eq:tPenetration}
\end{eqnarray}
To calculate $t\start$ for the primary population of ISN He we take $T_{\text{TOF}}$ for the spin angle bin with maximum flux for IBEX orbit 64 and $t_d$ corresponding to the actual observation time for this orbit $t_{\text{orb}}$. To account for all seasons of IBEX observations, we calculate $t\stopp$ using for $t_d$ a date increased by 10 years. The time of flight and its spread are calculated from Equations \ref{eq:TOFDefinition} and \ref{eq:TOFSpreadDefinition}, respectively. Consequently, for IBEX observations of ISN He, we have
\begin{eqnarray}
t\start & = & t_{\text{orb}} - T_{\text{TOF}} - \Delta \tau \nonumber \\
t\stopp & = & t_{\text{orb}} + 10\,\text{yr} - T_{\text{TOF}} + \Delta \tau, 
\end{eqnarray}
where $t_{\text{orb}}$ is the time for IBEX orbit 64 for the primary population and orbit 54 for the secondary population.

This can be done because for He atoms, which are insensitive to radiation pressure, the time of flight does not depend on the phase of the solar cycle. This interval in Figure~\ref{fig:sunspots} is marked in blue. To calculate the interval $(t\start, t\stopp)$ for the secondary population, we use the time of flight for the maximum flux bin for orbit 54. This interval is marked in pale blue in Figure~\ref{fig:sunspots}. 

For ISN H, we must differentiate between high and low solar activity. We make the estimates not for the full interval of IBEX observations but for two selected time moments: 2010.18 and for 2015.0. These dates are marked with orange and green dots, respectively, in Figure~\ref{fig:sunspots}. The penetration intervals are then calculated similarly as described in the preceding paragraph and marked in orange and green.

The penetration intervals for the primary and secondary population of ISN He are strikingly different. While IBEX observes the primary atoms that penetrated the interaction region 3--4 solar cycles before the observation, the secondary atoms that it observes were created in the outer heliosheath during an interval six solar cycles long, throughout most of the 19-th century. This suggests that the secondary He population brings information well averaged over the solar activity, and corresponding to time intervals that the only information on the solar activity we have available is from the sunspot record \citep[and possibly and indirectly, from early measurements of geomagnetic field variations,][]{nevanlinna:04a}. 

The penetration intervals for the primary ISN H vary with the solar activity level and in the length. They partly overlap with the penetration interval for the primary ISN He, but for solar minimum, they extend farther in time backwards. Effectively, the information brought by the primary atom to the detector is averaged over at least one solar cycle, with a delay of at least three solar cycles. 

\subsection{When will the atoms that crossed the heliopause during the Voyager OHS passage be observable at 1 au?}
\label{sec:VoyagerTimes}
\noindent
The Voyager spacecraft crossed the heliopause in August 2013 \citep[V1;][]{burlaga_etal:13a, stone_etal:13a, krimigis_etal:13a} and November 2018 \citep[V2;][]{richardson_etal:19a, stone_etal:19a, burlaga_etal:19b, krimigis_etal:19a}. An ability to sample at 1 au the ISN atoms that during this time were penetrating the interaction region would provide an opportunity to compare the data on the plasma state obtained in situ with effects of the OHS plasma-neutral interactions engraved in the ISN He and H populations. The question is when the atoms carrying this information will appear at 1 au.
\begin{figure}
\centering
\includegraphics[width=0.48\textwidth]{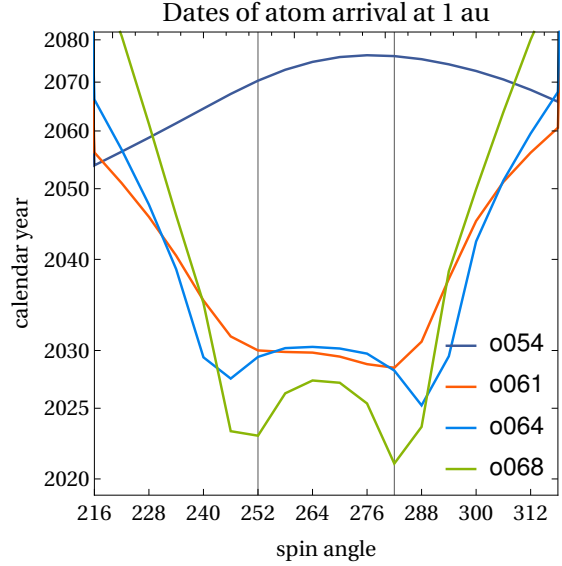}
\caption{Approximate arrival dates of ISN He atoms that were penetrating the heliopause at the time of heliopause crossing by Voyager 1, presented as a function of spin angle for IBEX viewing geometry characteristic for orbits 54, 61, 64, and 68. Note the vertical scale is logarithmic.}
\label{fig:heIBHPObsTimeV1}
\end{figure}

We approximately estimated these dates for ISN He adopting as a beginning of the visibility interval a date obtained as:
\begin{equation}
t_V = t_\text{V1,HP} +  T_{\text{TOF}}-\Delta \tau,
\label{eq:arrivalDateV1}
\end{equation}
where $T_{\text{TOF}}$ is the mean time of flight (Equation~\ref{eq:TOFDefinition}) of ISN He atoms from the heliopause at $r\HP$ to a detector at 1 au with the viewing geometry corresponding to that of IBEX, $t_{\text{V1,HP}}$ is the time of heliopause crossing by Voyager 1, and $\Delta \tau$ is the spread of times of flight (Equation~\ref{eq:TOFSpreadDefinition}). Equation~\ref{eq:arrivalDateV1} calculates calendar dates when the fastest of the ISN atoms that penetrated the heliopause after Voyager 1 crossed the heliopause will reach an IBEX-like detector at 1 au. These dates are shown in Figure~\ref{fig:heIBHPObsTimeV1}.

First of them, the fastest ones, are expected to be detectable by IBEX only at the end of the ISN observation season in 2022 (two pixels at the beginning of March). Most of them, however, will become observable approximately during the IMAP observation times \citep{mccomas_etal:18a}, between 2025 and 2030. Secondary atoms will be visible much later, well into the second half of the present century. The bulk of the atoms, both primary and secondary, will appear at Earth's orbit approximately one solar cycle after the first ones, beyond the expected lifetime of IMAP. 

For the primary H ISN, the first atoms that meet the Voayger at heliopause, will be seen at 1 au around 2027, and hopefully IMAP will be able to measure them. Some differences in times of flight from HP for hydrogen depending on the phase of the solar cycle exist on the order of 2 years, but they are not significant in our rough estimates. 

Note that interactions in the OHS that formed the secondary populations of ISN atoms and modified the primary ones must occur well before the heliopause crossing. Hence the conclusion from this part of the research is that currently operating or planned missions, during their lifetimes, most likely will not see the ISN atoms that were present in the interaction region when the Voyager spacecraft took their in-situ measurements of the plasma, magnetic field, and suprathermal particles.

\section{Summary and conclusions}
\label{sec:conclusions}
\noindent
We studied the time delay between penetration by ISN atoms of the region of intense plasma-neutral interaction in the OHS and detection of these atoms at 1 au. We showed that, as expected, this delay is much longer than that for the ENAs and varies from about 3---4 lengths of the solar cycle for the primary populations of ISN He and H, up to $\sim 150-200$ years for the secondary He, with a significant spread. For ISN H, the delay varies with the phase of the solar cycle. 

The length of the time of flight for the primary populations implies the atoms observed nowadays penetrated the interaction region in the OHS in 1960s to 1980s, i.e., during the space era, when the solar wind measurements are available at least for the ecliptic plane. This makes it potentially feasible to develop time-dependent models of the heliosphere with observation-based solar wind model to cover their OHS penetration times. However, at least for the secondary population of He there is no such possibility, as these atoms penetrated the interaction region in the 19-th century, when only sunspot observations are available. Therefore, it seems that analysis of ISN atom observations will have to rely on time-independent models with appropriately averaged solar wind conditions. 

Because of the different travel times between the primary ISN atoms, the secondary ISN atoms, and the ENAs observed at the same time, these observations bring information from very different epochs. Furthermore, the information brought by ISN atoms is averaged over several solar cycles. Therefore, it is not realistic to demand that simulation results performed using stationary model of the heliosphere will reproduce an instantaneous state of the heliosphere, characteristic for in-situ measurements such as those of energetic ions by the Voyagers. This also applies to the state of the heliosphere obtained from analysis of ENAs, especially those with the energies of tens of keV, which carry information from regions distant by hundreds au within several months. The bulk of ISN atoms that penetrated the heliopause at the time of Voyager crossing will become visible only at the end of and after the planned observation interval by IMAP, and the secondary population of He only in the second half of the present century.

The spread in the times of flight of the primary ISN He atoms on the order of the length of the solar cycle prevents an easy detection of hypothetical trends in the inflow direction of ISN gas to the heliosphere \citep{frisch_etal:13a}. The delay affects all observations of ISN He inside the heliosphere, including those by IBEX \citep{bzowski_etal:15a, schwadron_etal:15a, swaczyna_etal:18a} and by GAS/Ulysses \citep{witte:04, bzowski_etal:14a, wood_etal:15a, wood_etal:19a}. Results of these measurements were in agreement with each other within relatively large uncertainties, which was interpreted as no change. However, because of the spread in arrival times from the interstellar medium, the hypothetical change of a few degrees during $\sim 10$ years is unlikely to be detectable in these measurements. A much longer time interval of the observations is needed, on the order of three solar cycles. This will likely be available after the IMAP mission \citep{mccomas_etal:18b}, which is planned to terminate about 2030. This hypothetical change might only be detectable based on almost continuous observations of ISN He from Ulysses (1992--2007), IBEX and IMAP. 

\acknowledgments
{\emph{Acknowledgments}}. The sunspot time series was adopted from WDC-SILSO, Royal Observatory of Belgium, Brussels. This study was supported by Polish National Science Center grant 2015-19-B-ST9-01328.

\bibliographystyle{aasjournal}
\bibliography{iplbib}

\end{document}